\documentclass[twocolumn]{aa}
\usepackage{graphicx}
\usepackage[figuresright]{rotating}
\usepackage{txfonts}
\usepackage{booktabs}
\begin{document}
\title{A search for mid-infrared molecular hydrogen emission from protoplanetary disks
\thanks{Based on observations collected at the European Southern
Observatory, Chile (program ID 076.C-0129(A) and 078.C-0288(A))}}
\titlerunning{H$_2$ emission from protoplanetary disks}
   \author{A. Carmona
          \inst{1,2,}\thanks{{\it New address (from October 2007)}: ISDC, 
          chemin d'Ecogia 16, CH-1290 Versoix \& Geneva Observatory, University of Geneva, 
          chemin des Maillettes 51,
          CH-1290 Sauverny, Switzerland.}
          \and
          M.E. van den Ancker \inst{2}
          \and
          Th. Henning \inst{1}
          \and
          Ya. Pavlyuchenkov \inst{1}
          \and
          C.P. Dullemond \inst{1}
          \and
          M. Goto \inst{1}
          \and
          W.F-.Thi \inst{3}
          \and
          J.Bouwman \inst{1}
          \and 
          L.B.F.M. Waters \inst{4,5}
          }
   \offprints{A. Carmona\\
              \email{carmona@mpia.de}}             
   \institute{
              Max Planck Institute for Astronomy, K\"onigstuhl 17, 69117 Heidelberg, Germany
           \and  
              European Southern Observatory, 
              Karl Schwarzschild Strasse 2 , 85748 Garching bei M\"unchen, Germany
           \and
           Royal Observatory Edinburgh,
           Blackford Hill, Edinburgh, EH9 3HJ, UK
           \and
           Astronomical Institute, University of Amsterdam, 
           Kruislaan 403, NL-1098 SJ Amsterdam, The Netherlands,
           \and 
           Instituut voor Sterrenkunde, Katholieke Universiteit Leuven, 
           Celestijnenlaan 200B, B-3030 Heverlee, Belgium
           }
   \date{ }
   \abstract{
   We report on a sensitive search for mid-infrared molecular hydrogen emission from 
   protoplanetary disks.
   We observed the Herbig Ae/Be stars UX Ori, HD 34282, HD 100453,
   HD 101412, HD 104237 and HD 142666,
   and the T Tauri star HD 319139, and searched for H$\,_2~0-0~S(2)~(J=4-2)$ 
   emission at 12.278 micron and   
   H$\,_2~0-0~S(1)~(J=3-1)$ emission at 17.035 micron with VISIR, ESO-VLT's 
   high-resolution mid-infrared spectrograph.
   None of the sources present evidence for molecular hydrogen 
   emission at the wavelengths observed. 
   Stringent 3$\sigma$ upper limits to the integrated line fluxes and 
   the mass of optically thin warm gas ($T=$ 150, 300 and 1000 K) 
   in the disks are derived.
   The disks contain less than a few tenths of Jupiter mass of optically thin 
   H$_2$ gas at 150 K,
   and less than a few Earth masses of optically thin H$_2$ gas 
   at 300 K and higher temperatures.
   We compare our results to 
   a Chiang and Goldreich (1997, CG97) two-layer disk model of
   masses 0.02 M$_{\odot}$ and 0.11 M$_{\odot}$.
   The upper limits to the disk's optically thin warm gas mass 
   are smaller than the amount of warm gas
   in the interior layer of the disk,
   but they are much larger than the amount of molecular gas
   expected to be in the surface layer.
   If the two-layer approximation to the structure of the 
   disk is correct, 
   our non-detections are consistent with the low flux levels expected from the small amount 
   of H$_2$ gas in the surface layer.
   We present a calculation of the expected thermal H$_2$ emission from optically thick disks,
   assuming a CG97 disk structure, a gas-to-dust ratio of 100 and T$_{\rm gas}$ = T$_{\rm dust}$.
   We show that the expected H$_2$ thermal emission fluxes from 
   typical disks around Herbig Ae/Be stars
   are of the order of 10$^{-16}$ to 10$^{-17}$ erg s$^{-1}$ cm$^{-2}$ for a distance of 140 pc.
   This is much lower than the detection limits of our observations 
   (5 $\times$ $10^{-15}$ erg s$^{-1}$ cm$^{-2}$).
   H$_2$ emission levels are very sensitive to departures from 
   the thermal coupling between the molecular gas and dust in the surface layer. 
   Additional sources of heating 
   of gas in the disk's surface layer could have a major 
   impact on the expected H$_2$ disk emission.
   Our results suggest that in the observed sources the 
   molecular gas and dust in the surface layer have not significantly
   departed from thermal coupling (T$_{\rm gas}$/T$_{\rm dust}<$ 2) and 
   that the gas-to-dust ratio in the surface layer is very likely lower than 1000. 
   
   }
   \maketitle
%
%
%
\begin{table*}
\caption{Summary of the stellar properties and previous gas and dust observations. 
References: Acke et al. 2004 [A04], Acke et al. 2005 [A05], Dent et al. 2005 [D05],
de Zeeuw et al. 1999 [D99], 
Dunkin et al. 1997 [DU97], Hipparcos catalogue [HIP], Pi{\'e}tu et al. 2003 [P03],
van den Ancker et al. 1998 [V98], van der Plas et al. (in prep) [VP],
Stempels \& Gahm 2004 [S04].}             
\label{table:1}      
\centering                          
\begin{tabular}{l r c c c c c c c c l }        
\hline\hline                 
Star          & Sp.T.       & log($T_{eff}/{\rm K}$) & log($L$/L$_{\odot}$)  & V$_{\rm rad}$ & $d$  & Age$^a$   & CO sub-mm & $M_{DISK}^{\,\,\,\,\,\,\,\,\,\,\,\,\,b}$ & Group I/II $^c$ & References\\
              &             &       &    & [km s$^{-1}$]&  [pc] & [Myr] &           & [M$_J$]  &         & \\ 
\hline                        
UX Ori        & A4IVe     & 3.92         & 1.68  & 18      & 340    & 5        &  D05    & 16   & II& A04,D05,D99   \\
HD 34282      & A3Vne     & 3.94         & 1.27  & 16      & 400    & 7        &  D05    & 100  & I & A04,P03   \\
HD 100453     & A9Ve      & 3.87         & 0.90  & 17      & 112    & $>$ 10   &  ...    & 21   & I & A04,A05,HIP   \\ 
HD 101412     & B9.5Ve    & 4.02         & 1.40  & 17      & 160    & $>$ 8    &  ...    & ...  & II & A04,VP,D99    \\
HD 104237     & A4IVe+sh  & 3.92         & 1.54  & 13      & 116    & 2        &  ...    & 7.8   & II & A04,A05,V98   \\
HD 142666     & A8Ve      & 3.88         & 1.13  & 3       & 145    & 6        &  D05    & 16    & II & A04,DU97,D99  \\
HD 319139 $^d$ & K5Ve+K7Ve & 3.64+3.61   & 0.44  & -7      & 145    & \,\,\,2\,$^e$    &  ...       & \,\,\,\,\,47\,$^f$ & CTTS & S04,D99       \\
\hline
\end{tabular}
\flushleft
$^a$ Ages derived from the H-R diagram employing the pre-main-sequence
evolutionary tracks of Palla and Stahler (1993).
$^b$ Total disk gas masses are derived from dust continuum emission at millimeter wavelengths
assuming a gas-to-dust ratio of 100.
$^c$ Group of each Herbig Ae/Be star disk according to the classification scheme of Meeus et al. (2001).  
Group I sources -flared disks- are sources with a rising mid-IR spectral energy distribution (SED) and Group II sources -self-shadowed disks- are sources with a flat SED (Dullemond et al. 2002). CTTS: Classical T Tauri star.
$^d$ Spectroscopic binary of period 2.4 days [S04].
$^e$ We assume that both components have the same luminosity.
$^f$ Disk mass by Jensen et al. (1996) scaled to our adopted distance of 145 pc.
\end{table*}
\section{Introduction}

Circumstellar disks surrounding low- and intermediate-mass stars 
in their pre-main sequence phase are the locations where 
planets presumably form.
Such protoplanetary disks are composed of gas and dust.
Their mass is initially dominated by gas (99\%), 
specifically by molecular hydrogen (H$_2$),
which is the most abundant gas species.
The dust constitutes only  a minor fraction of the total disk mass,
however, it is the main source of opacity.
Consequently, 
most of what we know observationally about protoplanetary disks has been 
inferred from studies of dust emission and scattering  
(for recent reviews see Henning et al. 2006; Natta et al. 2007; Dullemond et al. 2007; Watson et al. 2007).
In order to understand the structure and evolution of protoplanetary disks, 
it is necessary to study their gaseous content independently from the dust.
For example, a basic physical quantity such as the disk mass
is conventionally deduced from dust continuum emission at millimeter wavelengths 
assuming an interstellar gas-to-dust ratio of 100 (e.g., Beckwith et al. 1990; Henning et al. 1994).
If dust is physically processed in the disk, as should be the case in order to form planets,
the gas-to-dust ratio must change with time.
The disk dissipation time scale 
- another fundamental quantity required to disentangle proposed giant planet formation scenarios 
(Pollack et al. 1996; Boss et al. 1998) -
is deduced from observations of thermal infrared excess emission produced by dust grains (Haisch et al. 2001, 2005).
Although recent studies (e.g., Sicilia-Aguilar et al. 2006) suggest a parallel evolution of the dusty and gaseous components, 
it still remains to be demonstrated that 
the gaseous disks disappear over the same time scale as the infrared excess. 
 
A variety of spectral diagnostics of the gas disk have been observed from the UV to the millimeter  
(see reviews by Najita et al. 2007; Dutrey et al. 2007).
However, the only diagnostic that is potentially able to probe 
the warm gas in the regions where giant planets are thought to form
is the mid-infrared (mid-IR) emission lines of H$_2$.
UV and near-infrared diagnostics only probe the innermost regions of the disk (R $<$ few AU),
and mm and sub-mm diagnostics are limited to probe the cold outermost regions of the disk (R $>$10 AU).
 
\begin{table*}
\caption{Summary of the observations.}             
\label{table:observations}      
\centering                          
\begin{tabular}{@{}l c l c c c c l c c c c c c c @{}}        
\hline\hline                 
Star    & $\lambda$ &  Date            &  U.T.    & $t_{exp}$ & Airmass $^a$ & V$_{\oplus\,{\rm rad}}\,^b$ & \multicolumn{2}{c}{Calibrator $^c$} & $t_{exp}$ & \multicolumn{2}{c}{Airmass} \\
        & [$\mu$m]  &                  &  [hh:mm] &   [s]        &              &  [km s$^{-1}$] & \multicolumn{2}{c}{ }          & [s]       & \multicolumn{2}{c}{       } \\

\hline                        
UX Ori    & 12.278 & 11 January 2006  & 02:28 & 3600 & 1.0 - 1.2  & 2.06
          & HD 36167  & (P)(F)  & 1000 & 1.1(P)  & 1.2(F) \\
          & 17.035  & 4 January 2007 & 01:54 & 3600 & 1.0 - 1.0  & 4.50
          & HD 25025 & (P)    & 600 &   1.0 (P)  & ...\\
HD 34282  & 12.278 & 10 January 2006 & 04:01 & 3600 & 1.0 - 1.5  & 2.17
          & HD 36167  & (P)(F)  & 1000 & 1.1(P)  & 1.7(F) \\
HD 100453 & 12.278  & 22 March 2006 & 07:27 & 3600 & 1.2 - 1.5  & 24.45
          & HD 89388  & (P)(F)  & 600   & 1.5 (P) & 2.0 (F) \\
          & 17.035  & 27 March 2006 & 06:26 & 3600 & 1.3 - 1.7  & 22.96
          & HD 89388  & (P)     & 1000 & 1.5 (P) & ... \\
HD 101412 & 12.278  & 30 March 2006 & 04:40 & 3600 & 1.2 - 1.4  & 23.82
          & HD 91056  & (P)(F)  & 600  & 1.3 (P) & 1.7 (F) \\
          & 17.035  & 30 March 2006 & 01:27 & 3600 & 1.3 - 1.2  & 24.02 
          & HD 91056  & (F)     & 1000 & ...     & 1.3 (F) \\
HD 104237 & 12.278  & 10 February 2006 & 05:43 & 3600 & 1.7 - 1.7  & 26.09
          & HD 92305  & (P)(F)  & 600  & 1.7 (P) & 1.8 (F) \\
          & 17.035  & 12 February 2006 & 06:36 & 3600 & 1.6 - 1.7  & 26.15
          & HD 92305  & (F)     & 1000 & ...     & 1.8 (F) \\
HD 142666 & 17.035  & 28 February 2006 & 06:28 & 3600 & 1.1 - 1.0  & 27.06
          & HD 169916 & (P)    & 1000 & 1.1 (P)  & ...\\
HD 319139 & 17.035  & 30 March 2006 & 08:30 & 3600 & 1.1 - 1.0  & 22.47
          & HD 169916 & (P)    & 1000 & 1.2 (P)  & ...\\
 
\hline  
\end{tabular}
\flushleft
$^a$ The airmass interval is given from the beginning to the end of the observations.
$^b$ V$_{\oplus\,{\rm rad}}$ is the expected velocity shift of the spectra due to the reflex motion of the Earth around the Sun and the radial velocity of the star.
$^c$ The standard stars were observed prior (P) and/or immediately following (F) the science observations.
\end{table*}

Molecular hydrogen is by far the most abundant molecular species in protoplanetary disks.
Unfortunately, H$_2$ is one of the most challenging molecules to detect.
Since H$_2$ is a homonuclear molecule, 
it lacks a permanent dipole moment and its transitions are thus electric quadrupole in nature.
The small Einstein coefficients, characteristic of the quadrupole transitions,
imply that H$_2$ emission features are very weak.
In addition, in the case of protoplanetary disks, 
the H$_2$ lines are not sensitive to the warm gas in the
optically thick regions where the dust and gas are at equal temperature. 
Practical observational challenges also have to be faced.
The mid-IR H$_2$ emission from the disk needs to be detected on the top of a strong mid-IR continuum.
From the ground, 
the mid-infrared windows are strongly affected by sky and instrument background emission,
and the H$_2$ transitions at 12 and 17 $\mu$m lie close to atmospheric absorption lines highly dependent on atmospheric conditions.
The advent of high spectral resolution spectrographs mounted on larger aperture telescopes,
allows for the first time the study of H$_2$ emission from the ground, 
but the search is still limited to bright targets.
From space, the problems of atmosphere absorption are alleviated and the $J = 2-0$ 
feature at 28 $\mu$m is visible. 
However, the beam sizes are large 
and the spectral resolution of space mid-infrared spectrographs are usually low when compared to ground-based facilities,
therefore, 
they are not very appropriate for small line-to-continuum ratios.

H$_2$ emission from protoplanetary disks in the mid-IR has been reported
from ISO observations (Thi et al. 2001). 
However, subsequent ground-based efforts (Richter et al. 2002; Sheret et al. 2003; Sako et al. 2005) 
did not confirm the ISO detections. 
H$_2$ emission in the mid-IR has been searched towards debris disks 
using Spitzer (Hollenbach et al. 2005, Pascucci et al. 2006,
Chen et al. 2006) with no detection reported. 
Most recently, 
Bitner et al. (2007) and Martin-Za\"idi et al. (2007) reported the detection of mid-IR H$_2$ emission in
two Herbig Ae/Be stars (AB Aur and HD 97048) from the ground, 
and Lahuis et al. (2007) reported the detection of
mid-IR H$_2$ emission in 6 T Tauri stars with Spitzer.

Here, we report on a sensitive search for molecular hydrogen emission from protoplanetary disks.
We observed six southern nearby (d$<$400 pc) Herbig Ae/Be   stars and one T Tauri star,  
employing the Very Large Telescope (VLT) imager and spectrometer for the  mid-infrared (VISIR; Lagage et al. 2004)\footnote{http://www.eso.org/instruments/visir}, 
ESO's VLT mid-infrared high-resolution spectrograph,
and searched for H$_2~0-0~S(1)~(J=3-1)$~emission at 17.035~$\mu$m and H$_2~0-0~S(2)~(J=4-2)$ emission at 12.278~$\mu$m.
The paper is organized as follows:
in Sect. 2 we present the sample studied, 
and the details of how the observations were performed.
In Sect. 3 we discuss the data reduction.
In Sect. 4 we deduce upper limits to the H$_2$ 
fluxes, and using the optically thin approximation, we 
derive upper limits to the mass of warm (150 - 1000 K) gas in the disks.
In Sect. 5 we discuss our results in the context of the
Chiang and Goldreich (1997) two-layer disk model. 
Finally, we summarize our results and conclusions in Sect. 6.
\begin{figure*}
\centering
\includegraphics[angle=0,width=0.95\textwidth]{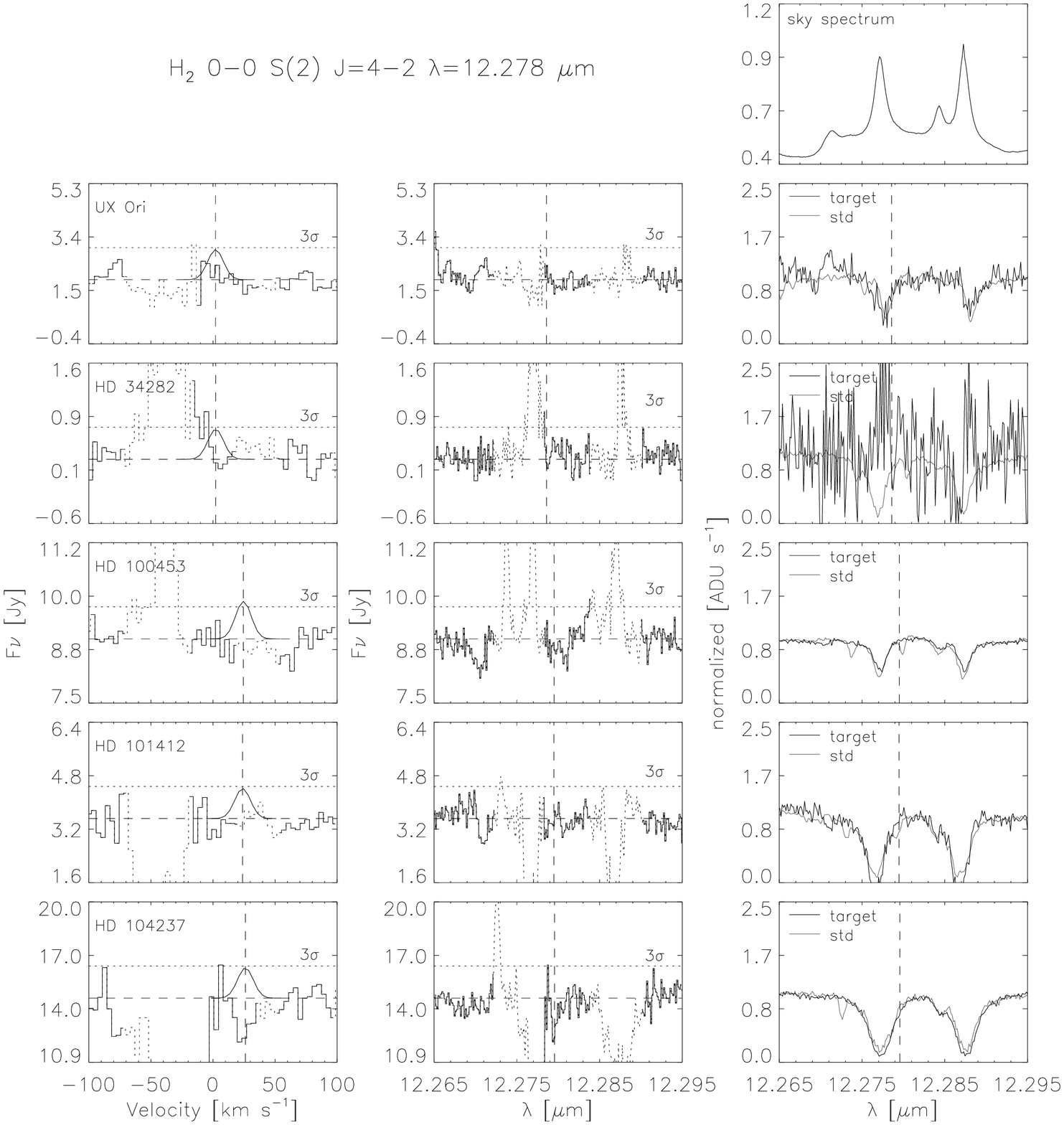}
      \caption{Spectra obtained for the H$_2$ 0--0 S(2) (J=4--2) line at 12.278 $\mu$m. 
               The left panel shows a zoom to the -100 to 100 km s$^{-1}$ interval of the atmospheric
               corrected spectra. 
               A Gaussian of {\it FWHM} = 15 km s$^{-1}$ and integrated line flux equal to
               the line-flux upper limits obtained is overplotted at the expected 
               velocity shifted location (vertical dashed line, see Table 2).
               The central panel shows the full corrected spectra. 
               Dotted lines show spectral regions strongly affected by telluric or standard star absorption features.
               The right panel shows the continuum normalized spectra of the standard
               star and the target before telluric correction.  
               The uppermost right panel displays the sky spectrum from a half-chop cycle.             
               The spectra are not corrected for the radial velocity of the targets.       }
\end{figure*}
\begin{figure*}
\centering
\includegraphics[angle=0,width=0.95\textwidth]{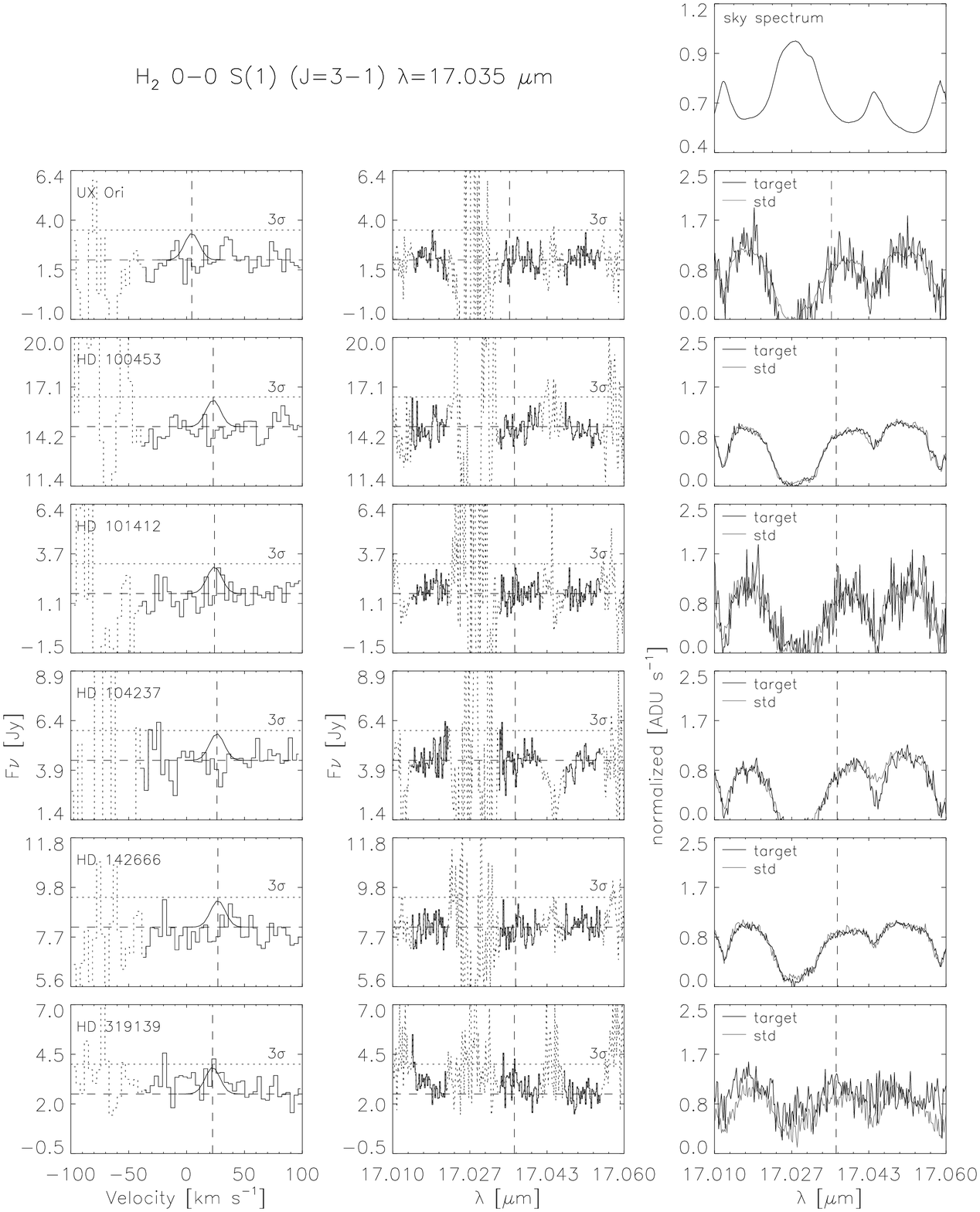}
      \caption{Spectra obtained for the H$_2$ 0--0 S(1) (J=3--1) line at 17.035 $\mu$m. 
               The left panel shows a zoom to the -100 to 100 km s$^{-1}$ interval of the atmospheric
               corrected spectra. 
               A Gaussian of {\it FWHM} = 15 km s$^{-1}$ and integrated line flux equal to
               the line-flux upper limits obtained is overplotted at the expected 
               velocity shifted location (vertical dashed line, see Table 2).
               The central panel shows the full corrected spectra. 
               Dotted lines show spectral regions strongly affected by telluric or standard star absorption features.
               The right panel shows the continuum normalized spectra of the standard
               star and the target before telluric correction.   
               The uppermost right panel displays the sky spectrum from a half-chop cycle.             
               The spectra are not corrected for the radial velocity of the targets.        
               }
\end{figure*}

\section{Observations}
{\it Target selection}.
We selected a sample of well known nearby Herbig Ae/Be   and T Tauri stars based on 
evidence of large disk reservoirs. 
The targets have either reported detections of cold CO gas at (sub)-mm wavelengths,
or dust continuum emission at mm wavelengths.
We chose stars with 12 $\mu$m continuum fluxes $>$ 0.5 Jy (otherwise too faint for acquisition with VISIR) 
and $<$ 25 Jy (hard to detect weak lines on top of a strong continuum). 
The physical properties of the targets are compiled in Table 1.

{\it Observations}.
The sources were observed in the first semester of 2006 and 2007
with VISIR,
a combined imager and spectrograph
designed for observations in the $N$ ($\approx 8-13~\mu$m)
and $Q$ bands ($\approx 16.5-24.5~\mu$m) (Lagage et al. 2004),
mounted at the ESO-VLT Melipal telescope in Cerro Paranal, Chile.  
The H$_2~0-0~S(1)~(J=3-1)$ line at 17.035 $\mu$m was observed 
in the {\it high-resolution long-slit mode} with a 0.4 arcsec slit,
giving a spectral 
resolution
R $\approx$ 21000, or 14 km s$^{-1}$.
The H$_2~0-0~S(2)~(J=4-2)$ line at 12.278 $\mu$m was observed 
in the {\it high-resolution echelle mode} with a 0.4 arcsec slit,
giving a spectral resolution R $\approx$ 20000, or 15 km s$^{-1}$.
   
The total integration time in each line was 1h.
The slit was oriented in the North-South direction.
Sky background was subtracted by chopping the telescope by $\sim$8" in the direction of the slit.
Asymmetrical thermal background of the telescope was subtracted by nodding the
telescope by $\sim$8" in the direction of the slit.
For correcting the spectra for telluric absorption and obtaining the absolute flux
calibration, spectroscopic standard stars at close airmasses to that of the science targets
were observed immediately preceding and following the 12~$\mu$m exposure, 
and preceding or following the 17~$\mu$m exposure\footnote{The 
12 $\mu$m window is much more affected by telluric absorption than the 17 $\mu$m
window. In order to be able to select the standard star observation
that provides the best telluric correction, observations of the H$_2$ S(2) line require the observation of a telluric standard 
preceding and
following the science observation.
The 17 micron window is less affected than the 12 micron window by telluric absorption, therefore, 
only one observation of a telluric standard (following or preceding the science observation) is required.}.
Finally, we tried as much as possible to observe the stars at a moment of 
high velocity shift between the telluric absorption feature at 12.27 microns and 
the H$_2$ S(2) line.
This effect is due to the reflex motion of the Earth around the Sun.
The typical PSF {\it FWHM} measured in the continuum was 0.5" ($\sim$4 pixels using the VISIR pixel scale of 0.127"/pixel)
equivalent to a spatial resolution of 50 AU for a source located at 100 pc. 
A summary of the observations is presented in Table 2. 

\section{Data reduction}
VISIR raw data consists of a collection of data cubes,
in which each data cube is associated with data taken in one nod  
position\footnote{The sequence of the observations is ABBA, where A and B correspond to two different nod positions.}.
The odd planes in the data cube correspond to a half-chop cycle frame,  
the even planes correspond to the average of the difference of all preceding half-chop cycles 
(i.e. A$_i$=A$_1$-B$_1$+~A$_2$-B$_2$+...+A$_i$-B$_i$/i).
The last plane corresponds to the average of all half-cycle difference images.

The first step in the data reduction process was to recover 
the individual half chop cycle frames from the data cube,
and to guarantee that all of them have the same wavelength 
in the same row of pixels (in each 2D spectrum, 
the spatial direction is along the columns and the dispersion direction is along the rows).
In each 2D half-chop cycle frame, 
at the position of the target spectrum, 
a cut with a size of 20 pixels in the spatial direction was made.
From this cut, a spectrum of the atmospheric emission was obtained
by the sum of the counts in the spatial direction (see uppermost right panel in Figs. 1 and 2).
This atmospheric emission spectrum was rebinned by a factor of 50 (i.e. 50 pixels where previously we had 1).
Using as reference the atmospheric emission spectrum from the first half-chop cycle of the first nod,
the offset that each half-chop cycle raw frame required for having the center
of the sky emission lines at the same pixel position
as the reference frame was determined.  
The offset was given by the difference between the center of the Gaussian
fitted to one sky emission line in the reference spectra,
and the center of the Gaussian fitted to the same sky line in the  
atmospheric emission spectrum of the half-chop 
cycle frame to be corrected.
The typical offsets were a fraction of a pixel (before rebinning),
and did not depend on the reference sky line.
Then, each 2D raw half-chop cycle frame was rebinned by a factor of 50 
(i.e. 50 pixels where previously we had 1) 
in the dispersion direction, 
and it was shifted -in the dispersion direction - by the offset found.
Finally, the obtained 2D frame was rebinned to the original size of 256x256 pixels. 
This procedure corrected small differences in 2D half-chop cycles introduced by
variations in the grating position within the instrument. 

Thereafter, with the corrected half-chop frames, 
the VISIR data cubes were re-built and processed with the VISIR pipeline (Lundin 2006).
Bad pixels were detected in the half-cycle frames when their number of counts exceeded the limit of 65000.
They were cleaned by interpolation with neighboring pixels. 
In the case of the 17 $\mu$m data (i.e. taken in the long-slit mode),
the 2D frames were corrected for distortion by interpolating the analytical optical distortion
(see description in the VISIR user's manual, Siebenmorgen et al. 2006) and the value from the source pixels.
The 2D frames in each nod position were created by averaging all the half-cycle difference images in the data cube.
The nodding images were produced by averaging the image in the two nod positions A and (-B), and dividing the result
by 2$\times$DIT, where DIT is the detector integration time\footnote{The factor 2 is due to the fact that the 
on-source and the off-source images both contribute to the whole DIT.}. 
The nodded images are jittered and were thus shifted and added to form the final combined image
using the offsets stored in the file header.

The wavelength calibration was performed with the following procedure.
The first half-cycle from the first nod position was ``collapsed",
producing a 1-dimensional spectrum of the atmosphere. 
The centers of the atmospheric emission lines (in pixels) were found by fitting Gaussians to their profiles
in a HITRAN model spectrum of Paranal's atmosphere emission (for details of the model, see Lundin 2006). 
The dispersion relation was found by a second degree polynomial fit between the centers of the
atmospheric lines in the observed and model spectrum.

The spectrum at each wavelength was extracted from the 2D 
combined image by summing the 
pixels in the spatial direction inside the PSF.
In the case of the long-slit 17~$\mu$m data, 
one spectrum was extracted for each of the nod positions present in the final 2D 
combined image.
The final extracted spectrum is the sum of these three spectra. 
The telluric correction was performed by dividing the extracted spectra by the extracted spectra of the standard star.
Small shifts (of a fraction of a pixel) on the spectra of the standard star in the wavelength 
direction were applied until the best signal to noise
spectra (i.e. best telluric correction) were found.
The spectral flux calibration was made by multiplying the telluric-corrected science spectrum with 
a model spectrum (Cohen et al. 1999) of the standard 
star\footnote{The catalogue of spectroscopic standard stars in the mid-IR is available at www.eso.org/VISIR/catalogue.}. 

In the case of the observations of the H$_2$ S(2) line in HD 34282, HD 100453 and HD 101412
the telluric standard stars showed photospheric absorption lines close to the location of the H$_2$ line.
In the case of the HD 34282 observation, the photospheric absorption is located outside 
the region where the H$_2$ line is expected (after correction by the velocity shift).
However, in the cases of HD 100453 and HD 101412 the expected location of the S(2) H$_2$ line 
partially overlaps with the location of the photospheric absorption feature in the 
spectrum of the standard star (see Fig. 3). 
This absorption feature is observed superimposed on the broad telluric absorption line at 12.277 micron. 
To avoid the error induced by this feature in the telluric corrected spectra, 
we removed this absorption feature from the standard star spectra 
by a Gaussian fit to the underlying broad atmospheric absorption feature at 12.277 micron (see Fig. 3).
After this correction, the science spectra were divided by the telluric standard and flux calibrated
in the conventional way.
\begin{figure}
\centering
\includegraphics[angle=0,width=0.5\textwidth]{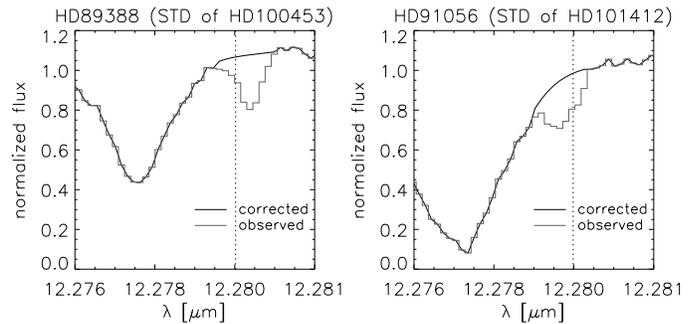}
      \caption{Correction for photospheric absorption features
      in the spectrum of the telluric standard (STD)
      of HD 100453 and HD 101412. In grey the observed
      STD spectrum, in black the corrected STD spectrum. The photospheric
      feature was removed by a Gaussian fit to the telluric line at 12.277 $\mu$m.
      The vertical dotted line displays the expected location of the S(2) H$_2$ line
      in the science spectrum. 
      }
\end{figure}
%
%
\begin{table*}
\caption{Upper flux limits.}             
\label{table:results}      
\centering                          
\begin{tabular}{@{} l c c c c c c c c c c c c c c c c c c c c c c c c c c c c c @{} }        
\hline\hline
\addlinespace[5pt]
          & \multicolumn{3}{c}{0--0 S(2) (J=4--2) 12.278~$\mu$m}         
          & \multicolumn{3}{c}{0--0 S(1) (J=3--1) 17.035~$\mu$m}\\[1mm]
\cmidrule(r){2-4}\cmidrule(r){5-7}
          & Continuum $^{a}$  & \multicolumn{2}{c}{LW = 15 km s$^{-1}$}      
          & Continuum $^{a}$  & \multicolumn{2}{c}{LW = 15 km s$^{-1}$}             \\ 
\cmidrule(r){3-4}\cmidrule(r){6-7}          
          & $F_{\nu}$  & Line Flux & Line Luminosity
          & $F_{\nu}$  & Line Flux & Line Luminosity \\[1mm]
 Star     & [Jy]       & [~$\times~10^{-14}$ ergs s$^{-1}$ cm$^{-2}$] & [~log\,$L/L_{\odot}$]               
          & [Jy]       & [~$\times~10^{-14}$ ergs s$^{-1}$ cm$^{-2}$] & [~log\,$L/L_{\odot}$]         \\[1mm]              
\hline                        
UX Ori    & 1.9 (1.1)  & $<$1.4 & $<$-4.3  
          & 2.0 (1.5)  & $<$1.3 & $<$-4.3 \\
HD 34282  & 0.3 (0.4)  & $<$0.5 & $<$-4.6  
          & ...        & ...    & ...     \\
HD 100453 & 9.0 (0.7)  & $<$0.9 & $<$-5.5
          & 14.8 (1.7) & $<$1.5 & $<$-5.2 \\
HD 101412 $^b$ & 3.5 (1.0)  & $<$1.2 & $<$-5.0
          & 1.6 (1.6)  & $<$1.4 & $<$-5.0 \\
HD 104237 & 14.6 (1.8) & $<$2.2 & $<$-5.0
          & 4.4 (1.5)  & $<$1.3 & $<$-5.3 \\
HD 142666 & ...        & ...    & ...
          & 8.0 (1.2)  & $<$1.1 & $<$-5.1 \\
HD 319139 & ...        & ...    & ...
          & 2.5 (1.5)  & $<$1.3 & $<$-5.1 \\
\hline                                   
\end{tabular}
\flushleft
$^a$ In brackets the 3$\sigma$ error in the continuum flux.
$^b$ Source also observed with Spitzer by Lahuis et al. (2007). They report a non-detection
of the H$_2$ S(1) and H$_2$ S(2) lines with 1$\sigma$ upper limits of $0.9\times10^{-14}$ erg s$^{-1}$ cm$^{-2}$ assuming a line width of
500 km s$^{-1}$.
\end{table*}

\begin{table*}
\caption{Upper warm H$_2$ mass limits.}             
\label{table:mass}      
\centering                          
\begin{tabular}{@{}l c c c c c c c c c@{} }        
\hline\hline                 
\addlinespace[5pt]
          & \multicolumn{6}{c}{Upper mass limits in $M_{J}\sim10^{-3} M_{\odot}$}     \\[1mm] 
\cmidrule(r){2-7}  
          & \multicolumn{3}{c}{12~$\mu$m}               
          & \multicolumn{3}{c}{17~$\mu$m} \\[1mm]
\cmidrule(r){2-4}
\cmidrule(r){5-7}
 Star     & 150 K       & 300 K  & 1000 K      
          & 150 K       & 300 K  & 1000 K     \\[1mm]              
\hline                        
\addlinespace[5pt]
UX Ori    & 27.9 & 1.9$\times 10^{-1}$ & 1.2$ \times 10^{-2}$
          & 1.0  & 6.8$\times 10^{-2}$ & 2.0$ \times 10^{-2}$\\
HD 34282  & 13.8 & 9.7$\times 10^{-2}$ & 6.0$ \times 10^{-3}$ 
          & ...   & ...   & ... \\
HD 100453 & 1.9 & 1.4$\times 10^{-2}$ & 0.9$ \times 10^{-3}$
          & 0.1 & 0.9$\times 10^{-2}$ & 2.5$ \times 10^{-3}$ \\
HD 101412 & 5.3 & 3.7$\times 10^{-2}$ & 2.3$ \times 10^{-3}$
          & 0.2 & 1.6$\times 10^{-2}$ & 4.8$ \times 10^{-3}$ \\
HD 104237 & 5.0 & 3.6$\times 10^{-2}$ & 2.2$ \times 10^{-3}$ 
          & 0.1 & 0.8$\times 10^{-2}$ & 2.3$ \times 10^{-3}$\\
HD 142666 & ... & ...   & ... 
          & 0.2 & 1.0$\times 10^{-2}$ & 3.1$ \times 10^{-3}$\\
HD 319139 & ...  & ...    & ...
          & 0.2 & 1.2$\times 10^{-2}$ & 3.7$ \times 10^{-3}$\\
\hline                                   
\multicolumn{6}{l}{Upper mass limits assuming optically thin H$_2$ in LTE.}
\end{tabular}
\end{table*}

\section{Results}
In Figs. 1 and 2 we present a compilation of the obtained spectra. 
In the left panel, the atmosphere-corrected flux calibrated spectra in wavelength scale 
measured by the velocities in km s$^{-1}$ are shown.
In the central panel, the full corrected spectra are presented.
In the right panel, for reference, the normalized spectra of the science target and the standard star are displayed.
None of the observed sources show evidence for H$_2$ emission at 12 or 17 $\mu$m.  

{\it Line flux upper limits}. 
In the flux-calibrated science spectra,
the standard deviation ($\sigma$) of the continuum flux was calculated in regions less influenced by telluric
absorption, photospheric features of the standard star and close to the features of interest.
In Figs. 1 and 2, the horizontal dotted line shows the 3$\sigma$ limits of the derived continuum flux.
3$\sigma$ upper limits to the integrated 
line flux were calculated by multiplying the 3$\sigma$ flux
with the instrument resolution line 
width\footnote{Here we assume a pole-on disk and that the true-line width is narrower 
than the instrumental line width of 15 km s$^{-1}$ (i.e. the spectral line is not resolved). 
If the line is broader, the sensitivity limits in Table 3 decrease proportionally with the {\it FWHM} (in km s$^{-1}$)
following $F=F_{15\,{\rm km}\,{\rm s}^{-1}}*({\rm {\it FWHM}}/15)$.}
of 15 km s$^{-1}$.
In Table 3, a summary of the upper flux limits is presented. 
In Figs. 1 and 2 a Gaussian line of {\it FWHM} = 15 km s$^{-1}$ and an integrated flux equal to the upper flux limit 
is overplotted on each observed spectrum.
The typical sensitivity limit of our observations is a line flux of 10$^{-14}$ ergs s$^{-1}$ cm$^{-2}$.
Our flux limits are of the order of magnitude of the 
H$_2$ S(1) and S(2) lines fluxes (1.1 and 0.53 $\times 10^{-14}$ ergs s$^{-1}$ cm$^{-2}$, respectively)
reported by Bitner et al. (2007) for AB Aur
and the H$_2$ line fluxes of 
0.33 - 1.70 $\times 10^{-14}$ ergs s$^{-1}$ cm$^{-2}$ reported for the H$_2$ S(2) line in the observations 
by Lahuis et al. (2007).  
Here, we should note that the observations of AB Aur were performed with TEXES, 
which provides a spectral resolution of 100000, increasing the line-to-continuum contrast compared
with our observations.

\begin{figure*}
\centering
\includegraphics[angle=0,width=0.53\textwidth]{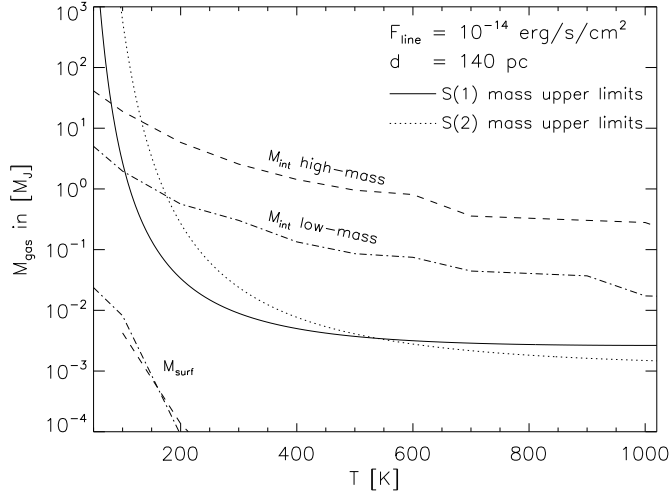}
      \caption{Mass limits of optically thin H$_2$ derived from H$_2$ S(1) (solid line)
      and H$_2$ S(2) (dotted line) as a function of the temperature for a 
      line flux limit of 10$^{-14}$ ergs s$^{-1}$ cm$^{-2}$ for a source 
      at a distance of 140 pc. 
      Dashed and dot-dashed lines show 
      the gas mass as function of the temperature for a Chiang and Goldreich (1997)
      optically thick two-layer model for a low-mass ($M_{\rm DISK}=0.02 M_{\odot}$)
      and a high-mass ($M_{\rm DISK}=0.11 M_{\odot}$) disk assuming a gas-to-dust ratio of 100. 
      $M_{\rm int}$ is the mass of the interior layer.
      $M_{\rm surf}$ is the mass of the surface layer.
      H$_2$ emission arises only from the optically thin molecular gas in surface layer of the disk.
      }
\end{figure*}

{\it Upper limits on the optically thin warm gas disk mass.}
Under the assumption that the H$_2$ emission and the emission of the
accompanying dust are optically thin,
that the emitting H$_2$ is in local thermodynamical equilibrium,
and that the source size is equal or smaller than VISIR's beam size,
we derived upper limits to the H$_2$ mass as a function of the temperature employing (Thi et al. 2001)

\begin{equation}
M_{{\rm gas}}= {f} \times 1.76 \times 10^{-20} \frac{4\pi d^{2}\,F_{ul}\,}{E_{ul}\,A_{ul}\,x_{u}(T)}\,\,\,M_{\odot},
\end{equation}
where $F_{ul}$ is the upper limit to the integrated line flux, 
$d$ is the distance in pc to the star,
$E_{ul}$ is the energy of the transition,
$A_{ul}$ is the Einstein coefficient of the $J=u-l$ transition\footnote{
$4.76 \times 10^{-10} {\rm\,s}^{-1}$ for the S(1) line $J=3-1$ transition,
$2.76 \times 10^{-9} {\rm\,s}^{-1}$ for the S(2) line $J=4-2$ transition 
(Wolniewicz, Simbotin, \& Dalgarno 1998).}
and 
$x_{u}$ is the population of the level $u$ at the excitation temperature $T$ in LTE;
$f$ is the conversion factor required for deriving the total gas mass from the
H$_2$-ortho or H$_2$-para mass determined. 
Since $M_{\rm H_2} = M_{\rm H_2\,ortho} + M_{\rm H_2\,para}$, $f$=1+ortho/para  for the S(2) line (a H$_2$-para 
transition) and $f$=1+1/(ortho/para) for the S(1) line (a H$_2$-ortho transition).
The equilibrium ortho-para ratio at the temperature $T$ was computed using the Eq. 1 of Takahashi (2001).  
In Table 4, we present our results. 

The H$_2$ S(1) and S(2) line fluxes
constrain the amount of optically thin hot and warm H$_2$ gas.
{\it The disks contain less than a few tenths of Jupiter mass of optically thin H$_2$ at 150 K,
and less than a few Earth masses of optically thin H$_2$ at 300 K and higher temperatures.}   
In Fig. 4 we present the mass of optically thin H$_2$ as a function of the temperature 
derived from a S(1) and a S(2) line with a flux $10^{-14}$ erg s$^{-1}$ cm$^{-2}$   
for a source at a distance of 140 pc.
This mass is the minimum amount of optically thin H$_2$ that would produce
a detectable H$_2$ line in our observations at a given temperature\footnote{We 
calculated a set of CG97 disk models around a prototypical Herbig Ae star (see Sect. 5, and 
parameters of MWC 480 in Table 5)
and found that the interior layer at $R>$ 5AU at 17 and 12 $\mu$m
would be optically thin if the mass of a disk with a size of 100 AU 
was smaller than 0.01 M$_J$,  
if a gas-to-dust ratio of 100 and T$_{\rm gas}$ = T$_{\rm dust}$ is assumed.}.
We see that the S(1) line sets constraints for the mass of optically thin H$_2$ at 
temperatures 150 K $<T<$ 500 K.
For higher temperatures the S(2) is a better tracer\footnote{For temperatures 
below 150 K the S(0) ($J=2-0$) line at 28 $\mu$m provides better constraints to the optically thin gas than the S(1)
line. However, as the Earth's atmosphere is completely opaque at these wavelengths,
the S(0) line is only observable from space.}.

\section{Discussion}

\subsection{Comparison with the gas mass of a two-layer disk model}
To understand the mass upper limits derived,
we suppose that the disk has a 
two-layer structure (Chiang \& Goldreich, 1997).
In this disk model,
the infrared excess in the SED is
produced by the radiation of small dust grains in the 
superheated optically thin surface layer of the disk, 
in addition to the emission of cooler dust in the mid-plane.
In the interior layer, 
the gas is thermalized by collisions with the dust grains,
the column densities of H$_2$ 
are larger than 10$^{23}$ cm$^{-2}$
and the optical depth is higher than one (see Fig. 6). 
The interior layer is optically thick in both gas and dust even at mid-infrared wavelengths.
Emission lines from an optically thick medium are only observable 
when there are temperature gradients. 
The hot upper layer can provide emission lines against the cold mid-plane.

Note that the two layer model is an approximation of the real structure of the disk.
In fact, in the uppermost zone of the surface layer, 
the gas is thermally decoupled from the dust, and 
due to intense X-ray or UV heating may well be ionized 
and at temperatures of the order of 10$^4$ K.
This uppermost surface is above the region where the hydrogen 
is molecular. 
The H$_2$ emission arises
from a layer below this uppermost surface at about A$_V$=1, or a hydrogen column
about 10$^{21}$ to 10$^{23}$ cm$^{-2}$ down from the surface, depending on the dust
opacity. For the remainder of the discussion, 
one must keep in mind
that when we speak about the disk surface layer, we 
refer specifically to the molecular gas layer of the disk's surface layer.

Using a two-layer disk model implementation (CGplus, Dullemond et al. 2001)
with physical parameters aimed to fit the SED of the prototypical Herbig Ae/Be
stars,  
we computed the expected amount of gas in the interior and surface layer as function
of the temperature for disks of mass 0.11 M$_{\odot}$
and 0.02 M$_{\odot}$ (see Table 5 and Fig. 5)\footnote{
In our models no puffed-up inner rim was used. 
We employed a model with a truncation radius at $T=$ 5000 K.
We assumed that T$_{gas}$=T$_{dust}$ and used a gas-to-dust ratio of 100. 
For disk-models with T$_{gas}$$\neq$T$_{dust}$ see for example,
Kamp and Dullemond (2004), Nomura and Millar (2005) and Jonkheid et al. (2006).}.
In Fig. 4 we present our results.
The mass limits derived from the H$_2$ S(1) and S(2) line observations
are smaller than the amount of warm gas in the interior layer, 
but much larger than the amount of molecular gas in the surface layer. 
Fig. 4 shows that the amount of gas in the surface layer is very small 
($< 10^{-2}$ M$_J \sim 3 $M$_{\oplus}$) 
and almost independent of the total mass of the disk.
{\it If the two-layer model is an adequate representation of the structure of the disk,
the thermal flux levels of H$_2$ mid-infrared emission are 
below the detection limit of the observations, 
because the mass of H$_2$ in the surface layer is very small.}

\begin{table}
\caption{Physical parameters of prototypical Herbig Ae/Be stars and their disks after Chiang et al. (2001).}             
\label{table:disk}      
\centering                          
\begin{tabular}{l l l}        
\hline\hline                 
             & MWC 480   & HD 36112   \\
   Parameter & high-mass & low-mass   \\
\hline
$T_{*}$ [K]                 & 8890  & 8465  \\    
$R_{*}$ [R$_{\odot}$]       & 2.1   & 2.1   \\
$M_{*}$ [M$_{\odot}$]       & 2.3   & 2.2   \\ 
$L_{*}$ [L$_{\odot}$]       & 24.6  & 20.2  \\
$d$ [pc]                    & 140   & 150   \\
$\Sigma_{0}$ [g cm$^{-2}$] & 8000  & 1000  \\
$a_0$ [AU]                  & 100   & 250   \\
$H/h$                       & 1.7   & 1.5   \\
$p$					        & 1.5   & 1.5   \\  	
$M_{\rm DISK}$ [M$_{\odot}$]& 0.11  & 0.02  \\
\hline                                   
\end{tabular}
\end{table} 
\begin{figure*}
\centering
\includegraphics[angle=0,width=\textwidth]{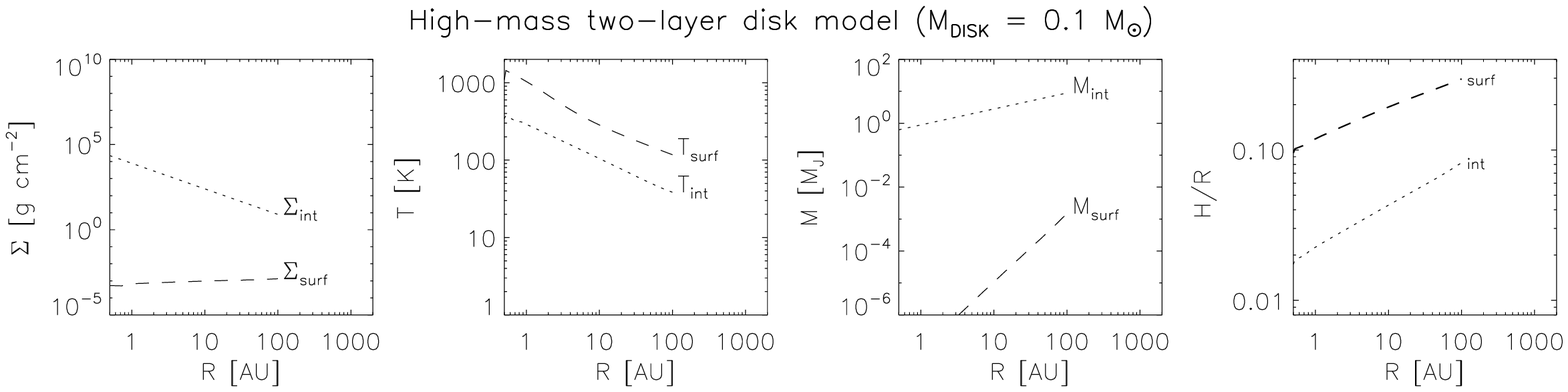}
\includegraphics[angle=0,width=\textwidth]{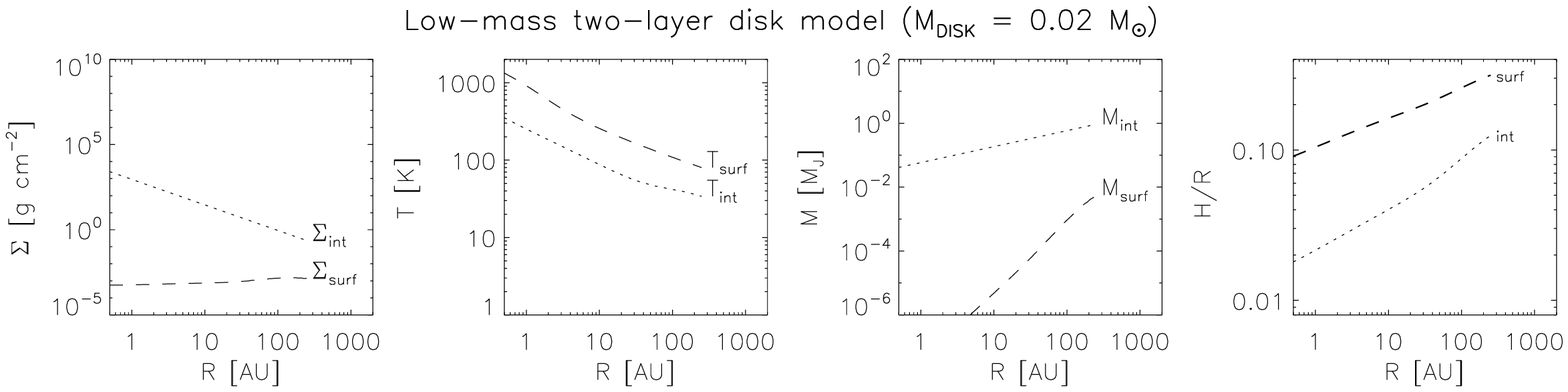} 
\caption{Physical parameters of the two-layer disk model as a function of radius (R). 
Upper panels: high-mass disk model ($M_{\rm DISK}=0.11M_{\odot}$, $L_{*}$= 24.6$L_{\odot}$). 
Lower panels: low-mass disk model ($M_{\rm DISK}=0.02M_{\odot}$, $L_{*}$= 20.2$L_{\odot}$).
The indices {\it int} and {\it surf} refer to the interior 
and the surface layer.
$\Sigma$ is the surface mass density,
T the temperature,
M the disk mass and H/R the scale height.
See Table 5 for details of the model parameters.}
\end{figure*}
\begin{figure*}
\centering
\includegraphics[angle=0,width=\textwidth]{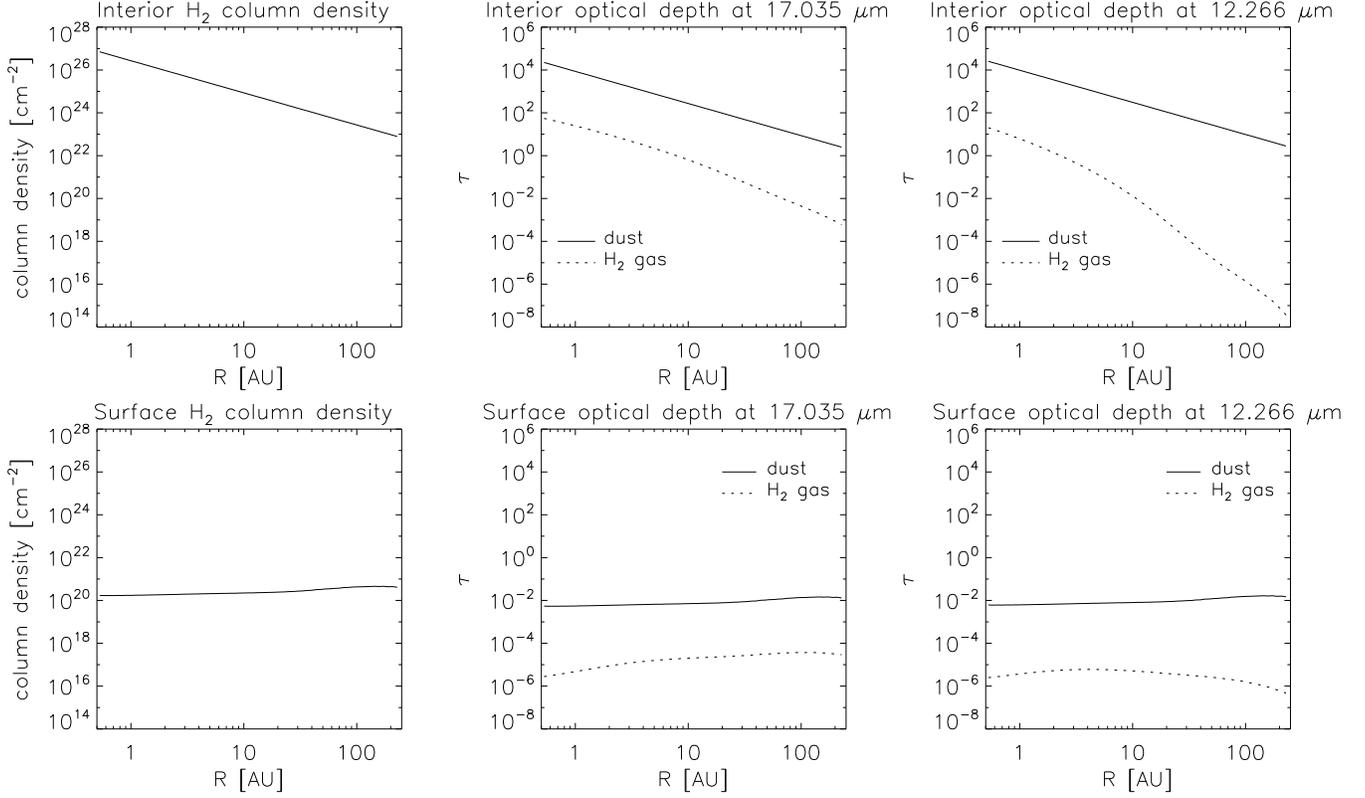}
      \caption{Column density of H$_2$ and optical depth of H$_2$ and dust
      as a function of radius $R$ for the interior (upper panels) 
      and the surface layer (lower panels) in the two-layer low-mass disk model 
      ($M_{\rm DISK}=0.02M_{\odot}$, $L_{*}$= 20.2$L_{\odot}$).
      A factor ten higher values in the H$_2$ column density and optical depth of the    
      interior layer are reached in the high-mass ($M_{\rm DISK}=0.11M_{\odot}$,  $L_{*}$= 24.6$L_{\odot}$) 
      model. See Table 5 for details of the model parameters.
      }
\end{figure*}


\begin{table*}
\begin{minipage}[t]{\textwidth}
\caption{Observational efforts to detect mid-IR H$_2$ emission from gas-rich protoplanetary disks.
References: Bitner et al. 2007 [B07], Carmona et al. (this work) [C07], Lahuis et al. 2007 [L07],
Martin-Za\"idi et al. 2007 [MZ07],
Richter et al. 2002 [R02], Sako et al. 2005 [S05], Sheret et al. 2003 [SR03], Thi et al. 2001 [T01].}
\label{table:disk}      
\centering                          
\renewcommand{\footnoterule}{}  
\begin{tabular}{l c c c c l l l}        
\hline\hline                 
Instrument   & $\lambda /\Delta \lambda $ at 17$\mu$m & LW            & beam         & flux \& flux limits &ref \\
             &             & [km s$^{-1}$]   & [arcsec $\times$ arcsec] &  [$\times$ 10$^{-14}$ erg s$^{-1}$ cm$^{-2}$] &  \\
\hline
VISIR        & 21000  & 14    & 0.4 $\times$ 0.4  & $<$0.5\,\,\,
			 & C07             \\
			 & ~~10000 \footnote{Observations of HD 97048 performed with a 0.75" slit, therefore, 
			 the different resolution.} &
			 30       & 0.4 $\times$ 0.4 & 2.4 &MZ07\\
COMICS 		 & 5000   & 60    & 0.6 $\times$ 0.4 & $<$1.0 
\footnote{The detection of H$_2$ S(1) in HD 163296, MWC863, CQ Tau and
LkCa15 by Thi et al. 2001 was not confirmed by Sako et al. 2005.}    & S05   \\
TEXES		 & 60000  & 5     & 2   $\times$ 2   & $<$1.4 
\footnote{The detection of H$_2$ S(1) in CQ Tau and AB Aur 
by Thi et al. 2001 was not confirmed by Richter et al. 2002.}      & R02   \\ 
             & 100000 & 3     & 0.4 $\times$ 0.4  & \,\,\,1.1
\footnote{Bitner et al. 2007 reported the detection of the H$_2$ S(1), S(2) and S(4) 
              lines in AB Aur.} 
			 & B07            \\
MICHELLE     & 15200  & 40    & 1   $\times$ 1   & $<$6.7 
\footnote{The detection of H$_2$ S(1) in HD 163296, AB Aur and GG Tau
by Thi et al. 2001 was not confirmed by Sheret et al. 2003.}      & SR03  \\
ISO-SWS      & 2000	  & 210   & 14  $\times$ 27  & $<$0.8 -- 40      $^{a,b,d}$& T01   \\
SPITZER
\footnote{From 76 sources studied one exhibits the H$_2$ S(1) line and six the H$_2$ S(2) line.
None of the 8 HAeBes observed exhibit H$_2$ mid-IR emission.}      &  600  & 500   &  5 $\times$ 5    & 0.32 
\footnote{Detection of the H$_2$ S(1) line in the T Tauri star Sz102.} & L07 \\  
\hline                                   
\end{tabular}
\end{minipage}
\end{table*} 

\subsection{Derivation of expected S(1) and S(2) H$_2$ emission from the two-layer disk model.}
Our observations are one of the most sensitive efforts to date to search for
H$_2$ emission from gas-rich protoplanetary disks (see Table 6).
If the H$_2$ emission is only produced from the upper layer of the disk,
as the two layer model suggests,
we would like to find out what flux levels we could expect
from typical Herbig Ae/Be disks, assuming T$_{\rm gas}$=T$_{\rm dust}$ and 
a gas-to-dust ratio equal to 100. For simplicity we first assume that the disk
is viewed pole-on ($i=0^{\circ}$). 
Here, we should note that the gas temperature
can be higher than the dust temperature in the surface layer 
(e.g., Kamp and Dullemond 2004, Nomura and Millar 2005, Jonkheid et al. 2006, Glassgold 2007).

In the two-layer disk model, at each disk radius 
the H$_2$ emission from the surface layer 
has to be observed on the top of the
optically thick continuum of the interior layer, 
and the optically thin continuum of the dust at the surface layer.
The total intensity (line plus continuum) emitted at each radius ($R$) is given by
\begin{equation}
I_{\nu}(R)=B_{\nu}(T_{\rm int})e^{-\tau_{\nu_{\rm surf}}}+ B_{\nu}(T_{\rm surf})(1-e^{-\tau_{\nu_{\rm surf}}}). 
\end{equation}
Here, $B_{\nu}(T_{\rm int})e^{-\tau_{\nu_{\rm surf}}}$ is the contribution of
the interior layer to the intensity.
$B_{\nu}(T_{\rm int})$ is the black body emission
of the interior layer at temperature $T_{\rm int}$ seen through the surface layer optical depth ($\tau_{\nu_{\rm surf}}=\tau_{\nu_{\rm dust\,surf}}+\tau_{\nu_{\rm gas\,surf}}$) . 
The contribution of the surface layer is $B_{\nu}(T_{\rm surf})(1-e^{-\tau_{\nu_{\rm surf}}})$.
It is the optically thin emission of the gas and dust in the surface layer at 
temperature $T_{\rm surf}$. 

The dust optical depth in the surface layer is given by
\begin{equation}
\tau_{\nu_{\rm dust\,surf}}=\kappa_{\nu_{\rm dust}} \Sigma_{\rm dust_{surf}},
\end{equation}
where $\kappa_{\nu_{\rm dust}}$ is the dust 
mass absorption coefficient\footnote{We used the opacities by Laor \& Draine (1993).} at the frequency $\nu$, 
and $\Sigma_{\rm dust_{surf}}$ is the dust surface mass density in the surface layer of the disk given by the CG97 model.
\begin{figure*}
\centering
\includegraphics[angle=0,width=\textwidth]{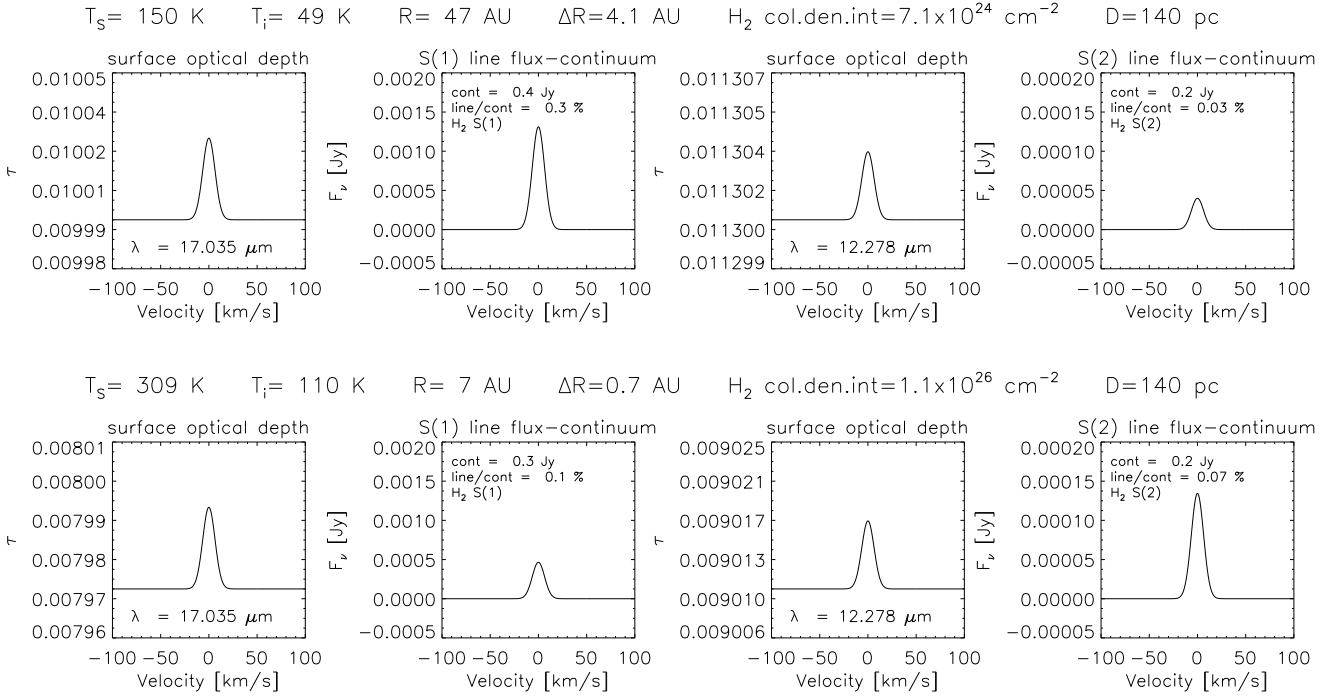}
      \caption{Surface optical depth and predicted H$_2$ S(1) and S(2) 
      emission line flux for the regions where the surface temperature ($T_s$)
      $\sim$ 150K (upper panel) and $\sim$ 300K (lower panel) 
      in the two-layer high-mass disk model ($M_{\rm DISK} = 0.11M_{\odot}$, $L_{*}$= 24.6$L_{\odot}$). 
      T$_i$ is the interior temperature. An instrumental line {\it FWHM} of 15 km s$^{-1}$ 
      was assumed. Note that the flux scale in the line flux-continuum 
      plot is 10 times smaller in the S(2) line plot as in the S(1) line plot.}
      
\end{figure*}
%
\begin{figure*}
\centering
\includegraphics[angle=0,width=\textwidth]{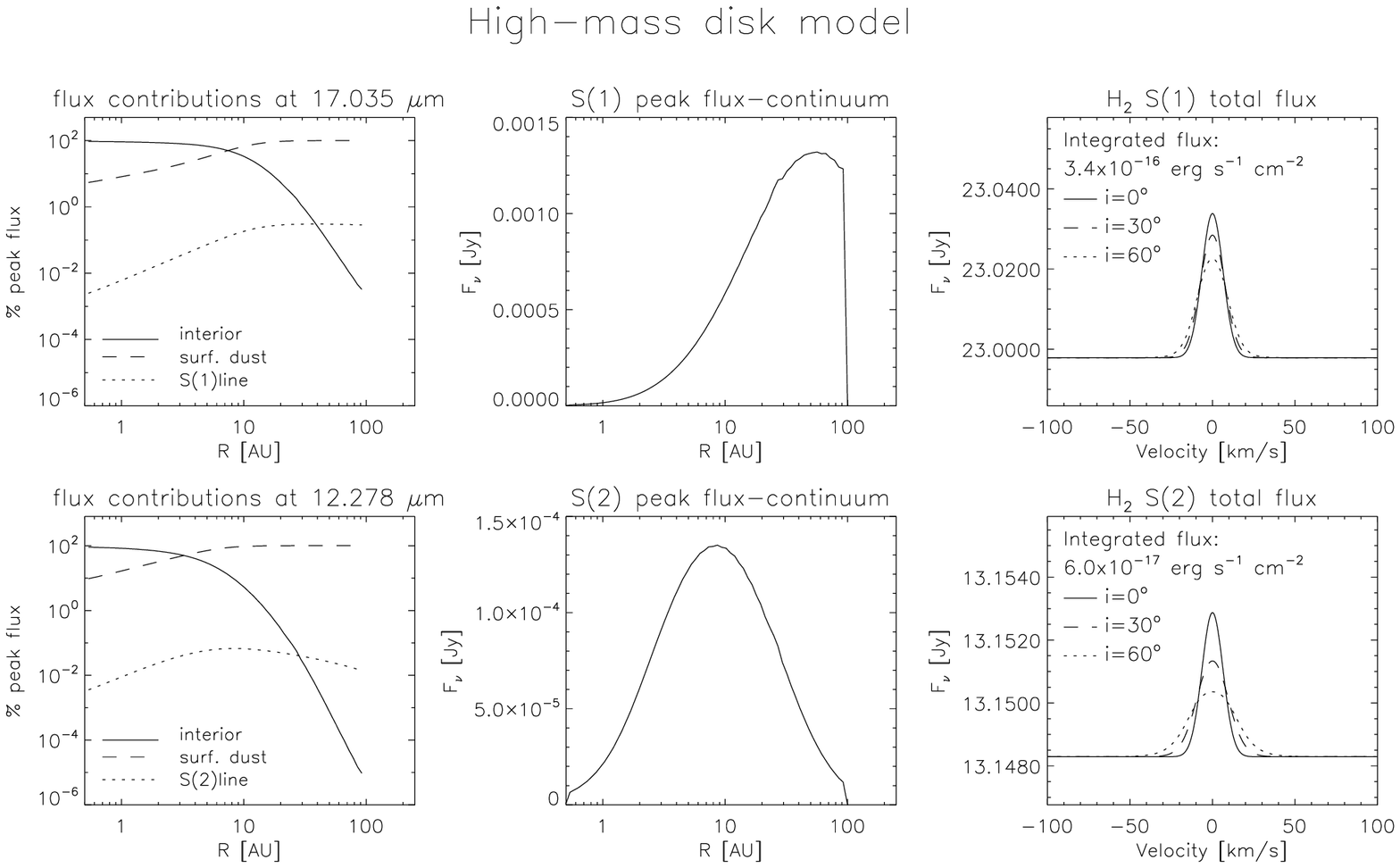}
\includegraphics[angle=0,width=\textwidth]{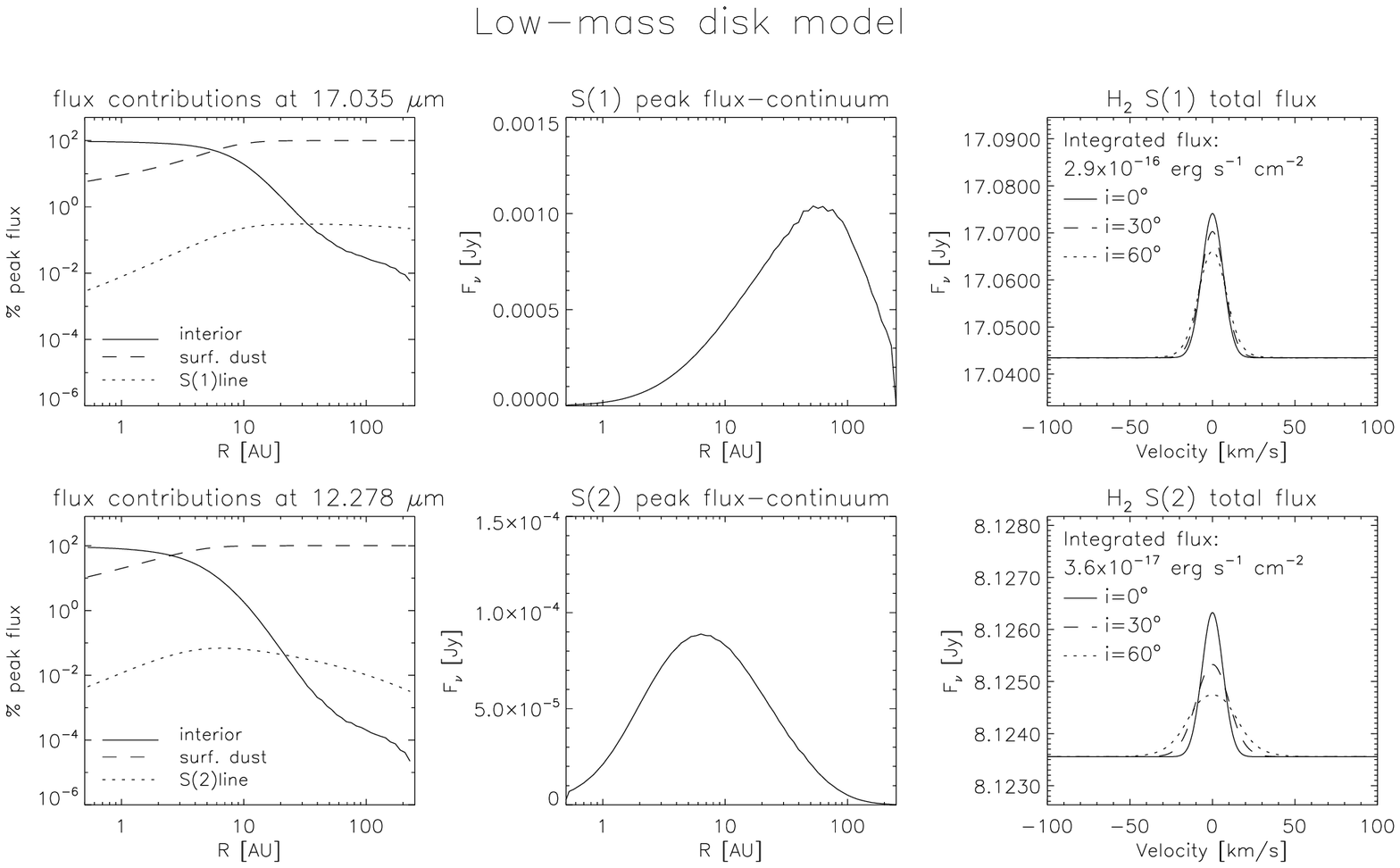}
      \caption{Total S(1) and S(2) line fluxes for the two-layer high-mass disk 
      model (upper six panels) and low-mass disk model (lower six panels). 
      Left panels present the percentage contribution to the peak flux by
      the optically thick interior layer (solid line), the optically thin 
      dust in the surface layer (dashed line) and the H$_2$ line (dotted line) as a function of radius
      for a pole-on disk ($i$=0$\,^{\rm o}$) .
      Central panels show the peak flux minus the continuum as a function of radius for 
      pole-on disk.
      Right panels display the expected total H$_2$ line flux for three disk inclinations:
      pole-on (solid line), $i$=30$\,^{\rm o}$ (dashed line)
      and $i$=60$\,^{\rm o}$ (dotted line). }
\end{figure*}

The H$_2$ gas optical depth in the surface layer at the frequencies $\nu$ 
in the vicinity of the H$_2$ line at frequency $\nu_0$ is 
\begin{equation}
\tau_{\nu _{\rm gas\,surf}}=\kappa_{\nu_{\rm line}}\Sigma_{\rm gas_{surf}}f_2,
\end{equation}
with $\Sigma_{\rm gas_{surf}}$ being the gas surface mass density in the disk surface 
layer (equal to 100 times $\Sigma_{\rm dust_{surf}}$)\footnote{Here we assume that all the gas is H$_2$.},
and $f_2$ the conversion factor from the total gas surface density 
$\Sigma_{\rm gas_{surf}}$ to the  
H$_2$-ortho or H$_2$-para surface densities. 
Since $M_{\rm gas} = M_{\rm ortho} + M_{\rm para}$, then $f_{2}$=1/(1+ortho/para)  for the S(2) line (a H$_2$-para transition) and $f_{2}$=1/(1+1/(ortho/para)) for the S(1) line (a H$_2$-ortho transition). The gas mass absorption coefficient ($\kappa_{\nu_{\rm line}}$)
in the vicinity of the H$_2$ transition at frequency $\nu_0$ is given by
\begin{equation}
\kappa_{\nu_{\rm line}}=\frac{1}{2m_{\rm p}}
\frac{h\nu_0}{4\pi}(x_l B_{lu}-x_u B_{ul}) \phi_{\nu}.
\end{equation}
Here, $2m_p$ is the mass of the H$_2$ molecule (2 times the mass of the proton $m_p$), 
$x_l$ and $x_u$ are the lower and upper level population of the
transition $J = u - l$, and $B_{lu}$ and $B_{ul}$ are the probabilities of absorption
and emission.

We assumed that the gas is in LTE and derived the level populations using
the Boltzmann equation.
The line profile $\phi_{\nu}$ is assumed Gaussian with a {\it FWHM}\footnote{We checked that the optical depth is smaller than one
with a $\phi$ with thermal width $\sigma_{th}=\sqrt{kT_s/(2m_p)}$, therefore, $\phi$ could be assumed from the start with a width equal to the instrument resolution.} 
equal to 15 km s$^{-1}$: 
\begin{equation}
\phi_{\nu}=\frac{1}{{\rm \sigma_{\nu}} \sqrt{\pi}}
e^{-\left(\frac{\nu - \nu_0}{\rm \sigma_{\nu}}\right)^2},
\end{equation}
where $\sigma_{\nu}= (\nu_0/c)*{\rm {\it FWHM}}/\left(2\sqrt{\rm ln2}\right)$.
The expected flux $dF_{\nu}(R)$  
at each radius (R) in the disk is given by
\begin{equation}
{\rm d}F_{\nu}(R)= I_{\nu}(R){\rm d}\Omega=I_{\nu}(R)\frac{2\pi R {\rm d}R}{d^2}
\end {equation}
with d$\Omega$ being the solid angle, and d$R$ the width of the region at 
a distance $R$ from the central
object where the surface temperature is equal
to $T_{\rm surf}$ and the interior temperature is equal to $T_{\rm int}$.
The distance to the source is $d$. 

Fig. 7 shows two examples of the surface optical depth ($\tau_{\nu_{\rm surf}}$) 
and the continuum subtracted S(1) and S(2) H$_2$ predicted line fluxes  
for the regions with surface temperature $\sim$150 K ($R \sim$47 AU upper panel) 
and $\sim$300 K ($R \sim$7 AU lower panel).
We assumed for the calculation 
a face-on high-mass disk model at a distance of 140 pc.
The column density of H$_2$ in  
the interior layer of the disk is larger than 10$^{23}$ cm$^{-2}$.
The total optical depth peaks at the 
H$_2$ S(1) and S(2) rest wavelengths.

The contribution to the total flux in each line
varies as a function of the distance to the central source.
Since at each radius, the surface layer optical depth of dust 
and H$_2$ gas is much smaller than one (see Fig. 6)
Eq. 2 can be approximated 
by
\begin{equation}
I_{\nu}(R)=B_{\nu}(T_{\rm int})+ 
B_{\nu}(T_{\rm surf})\tau_{\nu_{\rm dust\,surf}}+B_{\nu}(T_{\rm surf})\tau_{\nu_{\rm gas\,surf}}.
\end{equation}
In this way, the contributions to $I_\nu$ by the interior layer
$B_{\nu}(T_{\rm int})$, 
by the dust in the surface $B_{\nu}(T_{\rm surf})\,\tau_{\nu_{\rm dust\,surf}} $
and
by the H$_2$ line $B_{\nu}(T_{\rm surf})\,\tau_{\nu_{\rm gas\,surf}}$
at the surface can be separated at each radius. 
Combining Eq. 7 with Eq. 8, we can disentangle the different contributions to the
flux $dF_{\nu}$ as a function of the radius
\begin{equation}
dF_{\nu}(R) = dF_{\nu_{\rm\,int}}(R)+ dF_{\nu_{\rm\,dust\,surf}}(R) + dF_{\nu _{\rm\,gas\,surf}}(R).
\end{equation}

In the left panels of Fig. 8
we show the contributions, in percentage of the peak flux at each radius,
by the interior layer ($dF_{\nu_{\rm\,int}}(R)/dF_{\nu}(R)$, solid line), 
the dust in the surface ($dF_{\nu_{\rm\,dust\,surf}}(R)/dF_{\nu}(R)$, dashed lines) 
and the H$_2$ emission ($dF_{\nu_{\rm\,gas\,surf}}(R)/dF_{\nu}(R)$, dotted lines).
Up to a few AU, the main source of continuum emission is the dust in the interior layer;
at larger radii, 
the principal source of continuum emission is the superheated dust in the surface layer.
The H$_2$ line contribution is always very small, typically less than 1\% of
the total continuum flux.
     
In the central panels of Fig. 8
we show the evolution of line peak flux minus the continuum as a function of radius.
In the case of the S(1) line the maximum contribution to the total
flux is given by the region around 50-100 AU. 
In the case of the S(2) line the most important contribution to the line flux
is made by material around 5-20 AU.

The total expected flux from the disk is the sum of the contributions of each radius.
In the right panels of Fig. 8
we present the total expected line from the disk. 
The line strength over the continuum is less than 0.05 Jy
for the S(1) line and less than 0.005 Jy for the S(2) line.
The total integrated flux of the line is of the order of 10$^{-16}$
erg s$^{-1}$ cm$^{-2}$ for the S(1) line and 
10$^{-17}$ erg s$^{-1}$ cm$^{-2}$
for the S(2) line. 
{\it These line flux levels are two orders of magnitude 
below the sensitivity limits of our observations}.   

We note that our H$_2$ line flux predictions from the two-layer 
model only take into account thermal excitation of the gas, therefore,
show the minimum amount of expected flux.
If additional excitation mechanisms, such as X-rays or UV heating, 
are taken into account, the line fluxes are expected to be higher
because the gas is hotter. 
In sophisticated models of the disk
around the T Tauri star TW Hya (d = 56 pc) by Nomura et al. (2005, 2007), 
the predicted H$_2$ S(1) and S(2) line fluxes are of 
the order of a few $10^{-15}$ erg s$^{-1}$ cm$^{-2}$ and 
the line-to continuum flux ratios are approximately $10^{-5}$. 
For a star at 140 pc these line fluxes 
translate to a few $10^{-16}$ erg s$^{-1}$ cm$^{-2}$,
fluxes similar to our two-layer model\footnote{Note that T Tauri stars are less luminous than Herbig Ae/Be stars.
The fluxes are similar because the Nomura models include additional sources of heating beyond dust-gas collisions.}.
These theoretical H$_2$ line fluxes are still two orders of magnitude
below the detection limits of our observations. In addition, the 
achievement of line-to continuum flux ratios $<10^{-3}$ is a challenging task
for present ground-based observations. 
In our observations, for example, the best line/continuum ratio achieved is 10$^{-1}$. 

\subsubsection{The effect of the disk inclination} 
Until now, we assumed for simplicity that the disk is viewed pole-on. Now, 
if the disk is inclined by an angle $i$, 
the line profile $\phi_\nu$ used in Eq. 6
will be doppler-shifted for each parcel of gas that has a
velocity component in the line of sight.
For a parcel of gas located at radius $R$ and at azimuth\footnote{
We define $\theta$=0 as the angle where the velocity component of the motion is
0 in the line of sight.}
$\theta$, orbiting a star of mass $M_{\star}$,
the doppler shift $\Delta \nu$ due to the Keplerian motion of the gas is
\begin{equation} 
\Delta \nu=(\nu_{0}/c)~sin(\theta)sin(i)\sqrt{\frac{GM_{\star}}{R}},
\end{equation}
with $c$ being the speed of light. The doppler-shifted line profile $\phi_{\nu~{\rm shifted}}$
will be 
\begin{equation}
\phi_{\nu~{\rm shifted}}=\frac{1}{{\rm \sigma_{\nu}} \sqrt{\pi}}e^{-\left(\frac{\nu - \nu_0+\Delta\nu}{\rm \sigma_{\nu}}\right)^2}.
\end{equation}
Using Eq. 11 instead of Eq. 6 and employing Eqs. 3 to 8,
the intensity $I(R,\theta,i)$ and the flux $dF_{\nu}(R,\theta,i)$ emitted by each parcel of fluid of solid angle
$(R\,dR\,d\theta)/d^2$ is calculated.
The expected total emission by the disk is the sum of the contributions 
of all the fluid parcels from $R_{min}$ to $R_{max}$, 
the inner and  outer radius of the emitting region.
Assuming that $R_{min}$ = 0.5 AU and that $R_{max}$ is the disk outer radius, 
we calculated the expected H$_2$ S(1) and H$_2$ S(2) line profiles 
for the high-mass and the low-mass disk at inclinations 
of 30 and 60 degrees.
Our results are presented in the right panels of Fig. 8.
The net effect of the inclination is the broadening of the line and a decrease
of the peak flux. The S(2) line is more affected than the S(1) line by the 
inclination.
This is due to the fact that S(2) comes
from warmer gas, which is closer in and therefore orbiting faster.
Since most of the material responsible for the S(1) and S(2) 
emission is located at distances larger than a few AU (see central panels of
Fig. 8) even at high inclinations, at our spectral resolution, the double peaked profile is not exhibited.
This result suggests that higher temperature lines emitted in regions closer to the star
(e.g., the S(3) or S(4) lines) are more suitable for the investigation of inner disk dynamics.
\begin{figure*}
\centering
\includegraphics[angle=0,width=0.7\textwidth]{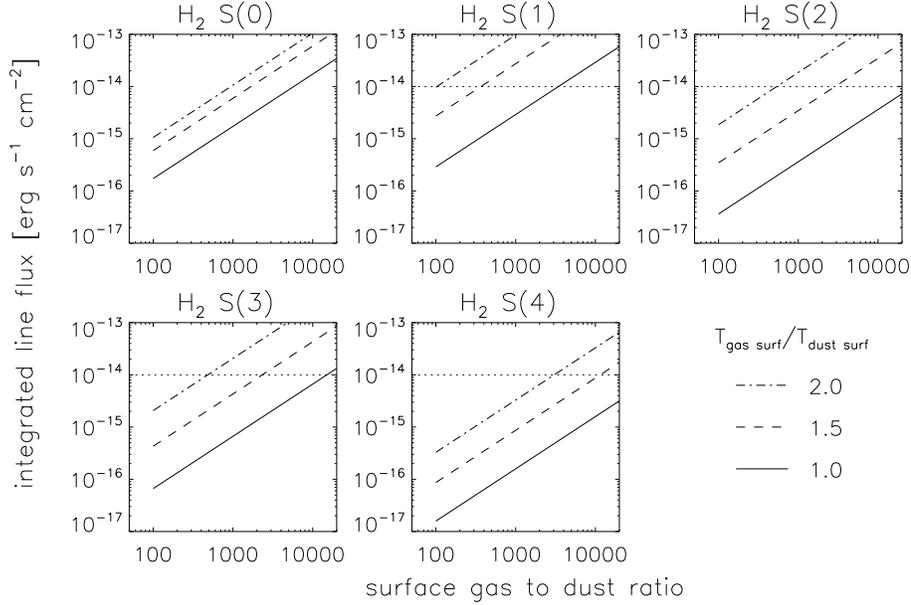}
      \caption{H$_2$ S(0), S(1), S(2), S(3), S(4) line fluxes for the two-layer low-mass disk 
      model as a function of the gas-to-dust ratio in the surface layer 
      for T$_{\rm gas}$/T$_{\rm dust}$ in the
surface ranging from 1.0 to 2.0. The dotted horizontal line presents typical ground detection limits. The S(0) line
is only observable from space.}
\end{figure*}
  
\subsection{Detections and non-detections of H$_2$ mid-IR emission from protoplanetary disks}
Our calculations of the expected thermal H$_2$ line fluxes from 
CG97 disks around typical Herbig Ae/Be stars, 
and sophisticated models of H$_2$ emission from disks around T Tauri stars 
by Nomura et al. (2005, 2007) 
show that the expected emission fluxes of H$_2$ from optically thick disks in the mid-IR are very weak 
($<10^{-16}$ erg s$^{-1}$ cm$^{-2}$ for a source at 140 pc).
These models provide a good explanation for the numerous non-detections
of H$_2$ mid-IR emission from optically thick disks reported so far
(GG Tau, HD 163296, Richter et al. 2002;
CQ Tau, Sheret et al. 2003; 
L551 IRS5, DG Tau, GW Ori, Richter et al. 2004;
HD 163296, MWC 863, CQ Tau and LkCa 15, Sako et al. 2005;
62 T Tauri and 8 Herbig Ae/Be stars, Lahuis et al. 2007;
UX Ori, HD 34282, HD 100453, HD 101412, HD 104237, HD 142666 and HD 319139, this work).

However, some detections of mid-IR H$_2$ emission from disks have been reported
in the Herbig Ae/Be stars AB Aur (Bitner et al. 2007)\footnote{H$_2$ S(1), S(2) and S(4) lines.} 
and HD 97048 (Martin-Za\"idi et al. 2007)\footnote{H$_2$ S(1) line.}, and in the T Tauri stars
Sz 102, EC 74, EC 82, Ced 110 IRS6, EC 92, and ISO-Cha 237 (Lahuis et al. 2007)\footnote{H$_2$ S(2) line, in addition the H$_2$ S(1) and S(3) lines in  Sz 102 and the H$_2$ S(3) line in EC 74.}.  
An interesting question to address is the reason for the high H$_2$ fluxes
observed in these sources.
Bitner et al. 2007, based on the gas temperature deduced  from the observed lines (670 K) 
and constraints 
on the line emitting region (18 AU) by a line profile fit, concluded, by comparison to 
a disk model of AB Aur (Dullemond et al. 2001), that in order to explain the observed emission,
an additional heating mechanism (X-rays or UV heating, e.g., Glassgold et al. 2007; Nomura et al. 2007) is needed to heat the gas in the disk upper layers (i.e. T$_{\rm gas\,surf}/$ T$_{\rm dust\,surf}>$ 1). 
A second possibility to explain high H$_2$ flux levels is to invoke  
a gas-to-dust ratio much larger than the canonical value of 100 in the surface layer 
of the disk (for example due to dust coagulation and sedimentation). 
Martin-Za\"idi et al. (2007) estimated the gas-to-dust ratio 
in HD 97048 from the detected H$_2$ S(1) line and the dust mass required for producing the observed flux level in the 
continuum. They obtained gas-to-dust ratios ranging from 3000 to 14000 for gas temperatures between 150 K to 1000 K. 

In order to explore the influence of a change in the gas-to-dust ratio and 
the thermal decoupling of gas and dust in the surface layer,
we calculated the expected S(0) to S(4) H$_2$ line fluxes from the low-mass disk model 
as a function of the surface gas-to-dust ratio for T$_{\rm gas\,surf}$/T$_{\rm dust\,surf}$
ranging from 1.0 to 2.0. 
We proceeded to use a simplified prescription. 
The output of a CG-plus model was taken as the disk surface mass density and temperature of the dust
in the upper and interior layer. 
To obtain the molecular gas surface mass density in the upper layer, 
the dust surface mass density in this layer was multiplied by the gas-to-dust ratio.
To obtain the molecular gas temperature  in the
upper layer, the dust temperature in this layer was multiplied by T$_{\rm gas\,surf}$/T$_{\rm dust\,surf}$.
Our results are presented in Fig. 9. 

As expected, an increase in the gas-to-dust ratio or T$_{\rm gas\,surf}$/T$_{\rm dust\,surf}$
yields an increase in the H$_2$ flux emission levels. 
However, {\it deviations from T$_{\rm gas\,surf}$=T$_{\rm dust\,surf}$ have 
a much larger impact on the increase of the strength of the H$_2$ lines than 
the changes in the gas-to-dust ratio}. 
For example, to obtain flux levels over $10^{-14}$ erg s$^{-1}$ cm$^{-2}$ in the S(1) 
line 
- sufficient to explain the detections by Bitner et al. and Mantin-Za\"idi et al. - 
it is just required that T$_{\rm gas\,surf}$/T$_{\rm dust\,surf}>$ 2.0. 
In contrast, to achieve similar S(1) flux levels the gas-to-dust ratio should increase by 
a factor of thirty. The most likely scenario is that in the molecular upper layer of the disk,
both the gas temperature and the gas-to-dust ratio change.
For example,
a simultaneous increase in the gas-to-dust ratio ($>$ 1000) and gas temperature
(T$_{\rm gas\,surf}$/T$_{\rm dust\,surf}>$ 2.0)
can boost the flux levels of the H$_2$ S(2) line to detectable levels.
Therefore, the non detections in our observations suggest that in the
observed disks, the molecular gas and dust in the surface layer has not significantly departed from
thermal coupling and that the gas-to-dust ratio in the surface layer is not larger than 1000.

One interesting aspect is that observations from space (Lahuis et al. 2007)
report more numerous detections of the S(2) line than detections of the S(1) line, although
in our LTE models the S(1) line is in general stronger than the S(2) line.
Note that for a given temperature the amount of ortho H$_2$ (responsible for the S(1) line)
is higher than the amount of para H$_2$ (responsible of the S(2) line).    
This discrepancy suggests that the level populations and the
ortho/para ratio of the H$_2$ producing the line in the observed sources are not in LTE. 
One additional piece of information 
suggesting that the H$_2$ gas responsible for the line emission
is heated by a mechanism other than collisions with dust, 
is the detection of a H$_2$ S(4) line 50\% 
stronger than the H$_2$ S(1) line in AB Aur (Bitner et al. 2007).
In our LTE disk models the H$_2$ S(4) line is always much weaker than the H$_2$ S(1) line.
Our results confirm the suggestion by Bitner et al. (2007) that in AB Aur 
the H$_2$ emission is produced in a narrow region (at 18 AU),
in which the gas has been excited to a high temperature (T $\simeq$670K).
We speculate that this could be the location of shock-heated gas in the AB Aur disk.

\section{Summary and conclusions}

We observed a sample of nearby pre-main sequence stars with evidence 
for cold ($T < 50$ K) 
gas disk reservoirs and searched for emission of the warm gas with $T > 150$ K, 
which is expected to be present in the inner region of these disks.
None of the targets show any evidence for H$_2$ emission at 17.035 $\mu$m 
or at 12.278 $\mu$m.
From the 3$\sigma$ upper limits of the H$_2$ line fluxes, 
we found stringent upper limits to the mass of optically thin warm H$_2$.
The disks contain less than a few tenths of Jupiter mass of optically thin H$_2$ at 150 K,
and less than a few Earth masses of optically thin H$_2$ at 300 K and higher temperatures. 

Assuming that T$_{\rm gas}$=T$_{\rm dust}$ and a gas-to-dust ratio of 100,
we compared our results to models of disks employing 
a Chiang and Goldreich (1997) optically thick two-layer disk model of
masses 0.02 M$_{\odot}$ and 0.11 M$_{\odot}$.
The upper limits to the disk optically thin warm gas mass are smaller 
than the warm gas mass 
in the interior layer of the disk,
but they are much larger than the amount of molecular gas expected to be in the surface layer.
The amount of mass in the surface layer is very small ($<$ 10$^{-2}$ M$_J$)
and almost independent of the total disk mass.
We calculated the expected H$_2$ S(1) and H$_2$ S(2) line fluxes
emitted from a two-layer disk for the low-mass  
and the high-mass cases assuming a distance of 140 pc,
and LTE thermal emission.
The predicted line fluxes of the two-layer disk model
are of the order of $\sim$10$^{-16}-10^{-17}$ erg s$^{-1}$ cm$^{-2}$,  
much smaller than the detection limits of our observations 
(5 $\times$ $10^{-15}$ erg s$^{-1}$ cm$^{-2}$).

{\it
If the two-layer approximation to the structure of the 
disk is correct, 
we are essentially ``blind" to most of the warm H$_2$ in the disk
because it is located in the optically thick interior layer of the disk.
Our non-detections are explained because of the small flux levels expected from 
the little mass of H$_2$ present in the optically thin surface layer.}
Naturally, the two-layer disk model is only an approximation of the
real structure of a protoplanetary disk. The puffed-up inner rim,
which could be an important contributor to the H$_2$ emission, 
is not included in our models, and in reality there will be
a smooth transition zone between the disk hot surface layer and the 
cool disk interior, which again could contribute significantly to the H$_2$ 
emission. In addition, the surface layers of the disk are 
likely to have gas temperatures hotter than the dust temperatures
or a gas-to-dust ratio higher than the canonical value of 100.
Both effects could potentially increase the H$_2$ 
emission by significant amounts. 
We presented additional calculations in which the gas-to-dust ratio and the temperature of the
molecular gas in the surface layer of the disk were increased. 
We showed that detectable S(1) and S(2) H$_2$ line flux levels 
can be achieved if T$_{\rm gas\,surf}$/T$_{\rm dust\,surf}>$ 2 and if the  gas-to-dust ratio in the surface 
layer is greater than 1000. 
H$_2$ emission levels are very sensitive to departures from 
the thermal coupling between the molecular gas and dust in the surface layer. 
Our results suggest that in the observed sources the molecular gas and the dust in the surface layer have not significantly
departed from thermal coupling and that the gas-to-dust ratio in the surface layer is very likely lower than 1000. 
A definitive interpretation of our results
awaits the development of future, more sophisticated models.

\begin{acknowledgements}
This research has made use of the SIMBAD database
operated at CDS, Strasbourg, France.
We would like to thank the staff of Paranal Observatory for performing our observations
in service mode.
We also would like to thank the referees, whose comments have helped 
to improve the content and presentation of our paper.
\end{acknowledgements}

\end{document}